\documentstyle[eqsecnum,aps,preprint,graphicx]{revtex}

\makeatletter

\def\compoundrel#1\over#2{\mathpalette\compoundreL{{#1}\over{#2}}}
\def\compoundreL#1#2{\compoundREL#1#2}
\def\compoundREL#1#2\over#3{\mathrel
  {\vcenter{\hbox{$\m@th\buildrel{#1#2}\over{#1#3}$}}}}
\makeatother

\newcommand{\mm}{\langle m_\nu \rangle}
\newcommand{\mmmax}{\langle m_\nu \rangle_{\mbox{\tiny max}}}

\begin{document}
\def\btt#1{{\tt$\backslash$#1}}
\draft
\title{{\bf CP Violations in Lepton Number 
Violation Processes and Neutrino Oscillations}}
\author{K. MATSUDA, N. TAKEDA and T. FUKUYAMA}
\address{Department of Physics, \\
        Ritsumeikan University, Kusatsu, \\
        Shiga 525-8577, Japan}
\author{H. NISHIURA}
\address{Department of General Education, \\
        Junior College of Osaka Institute of Technology, \\
        Asahi-ku, Osaka 535-8585, Japan}

\date{March 4, 2000}
\maketitle



\begin{abstract}
We examine the constraints on the MNS lepton mixing matrix from the 
present and future experimental data of the neutrino oscillation and 
lepton number violation processes.
We introduce a graphical representation of the $CP$ violation phases 
which appear in the lepton number violation processes such as neutrinoless 
double beta decay, the $\mu^--e^+$ conversion, and the K decay, 
$K^-\to\pi^+\mu^-\mu^-.$ Using this graphical representation, we 
derive the constraints on the $CP$ violation phases 
in the lepton sector.
\end{abstract}

\section{Introduction}
From the recent neutrino oscillation experiments it becomes affirmative 
that neutrinos have masses. The present and near future experiments enter 
into 
the stage of precision tests for masses and lepton mixing angles. In 
this situation 
the investigation of the $CP$ violation effects in the lepton sector has 
become 
more and more important. By taking account of the possible leptonic $CP$ 
violating phases for Majorana neutrinos, 
we have obtained the constraints on the lepton mixing angles from the 
neutrinoless double beta decay 
($(\beta\beta)_{0\nu}$)\cite{fuku}\cite{fuku2},  
the \(\mu^-\)-\(e^+\) conversion and the K decay, 
$K^- \rightarrow \pi^+\mu^-\mu^-$\cite{nishi}.
In this paper, we propose graphical representations of the 
$CP$ violating phases which appear in those lepton number 
violating processes. By using those representations, 
we derive the allowed regions on the leptonic $CP$ violating phases from 
$(\beta\beta)_{0\nu}$ without using any constraints on the mixing angles.
We also try to determine the magnitude of the $CP$ violating phases 
by combining the constraints on the neutrino masses and 
mixing matrix elements from  
the recent Super Kamiokande atmospheric neutrino experiment 
\cite{skamioka}, 
the solar neutrino experiment 
\cite{kamioka}\cite{homestake}\cite{gallex}\cite{sage}\cite{fogli} , 
the recent CHOOZ reactor 
experiment \cite{chooz} , and the future KamLAND reactor 
experiment \cite{kamland} with those from the lepton number violating 
processes such as $(\beta\beta)_{0\nu}$.
\par  
The amplitudes of those three lepton number violating processes are, in 
the absence of 
right-handed weak couplings, proportional to the "averaged" masses 
$\langle m_{\nu} \rangle_{e e}$, $\langle m_{\nu} \rangle_{\mu e}$ 
and $\langle m_{\nu} \rangle_{\mu \mu}$ . The "averaged" mass  
$\langle m_{\nu} \rangle_{e e}$ defined from 
$(\beta\beta)_{0\nu}$ is given \cite{doi} by 
\begin{equation}
\langle m_{\nu} \rangle_{e e}  = |\sum _{j=1}^{3}U_{ej}^2m_j| . 
\label{eq761}
\end{equation}
Similarly, the "averaged" masses  $\langle m_{\nu} \rangle_{\mu e}$ 
defined from 
\(\mu^-\)-\(e^+\) conversion and $\langle m_{\nu} \rangle_{\mu \mu}$
defined from the lepton number violating K decay,
$K^- \rightarrow \pi^+\mu^-\mu^-$ are given \cite{takasugi} 
\cite{nishiura} by

\begin{eqnarray}
\langle m_{\nu} \rangle_{\mu e} & =& |\sum _{j=1}^{3}U_{\mu j}U_{e 
j}m_j| , \label{eq763}\\
\langle m_{\nu} \rangle_{\mu \mu} & =& |\sum _{j=1}^{3}U_{\mu j}^2m_j| , 
\label{eq771}
\end{eqnarray}
respectively.
The $CP$ violating effects are included in the "averaged" masses 
$\langle m_{\nu} \rangle_{e e}$, $\langle m_{\nu} \rangle_{\mu e}$ 
and $\langle m_{\nu} \rangle_{\mu \mu}$ defined in Eqs.(\ref{eq761}) 
$\sim$ 
(\ref{eq771}).
Here $U_{a j}$ is the Maki-Nakagawa-Sakata (MNS) left-handed lepton 
mixing 
matrix which combines the weak eigenstate neutrino ($a=e,\mu$ and $\tau$) 
to the mass eigenstate neutrino with mass $m_j$ ($j$=1,2 and 3).  
The $U$ takes the following form in the standard representation 
\cite{fuku}:
\begin{equation}
U=
\left(
\begin{array}{ccc}
c_1c_3&s_1c_3e^{i\beta}&s_3e^{i(\rho-\phi )}\\
(-s_1c_2-c_1s_2s_3e^{i\phi})e^{-i\beta}&
c_1c_2-s_1s_2s_3e^{i\phi}&s_2c_3e^{i(\rho-\beta )}\\
(s_1s_2-c_1c_2s_3e^{i\phi})e^{-i\rho}&
(-c_1s_2-s_1c_2s_3e^{i\phi})e^{-i(\rho-\beta )}&c_2c_3\\
\end{array}
\right).\label{eq772}
\end{equation}
Here $c_j=\cos\theta_j$, $s_j=\sin\theta_j$ 
($\theta_1=\theta_{12},~\theta_2=\theta_{23},~\theta_3=\theta_{31}$). 
Three 
$CP$ violating phases, $\beta$ , $\rho$ and $\phi$ appear in $U$ for 
Majorana
neutrinos \cite{bilenky}. In this paper we introduce the graphical 
representations of 
the complex masses $\sum _{j=1}^{3}U_{ej}^2m_j, \sum _{j=1}^{3}U_{\mu j}
U_{e j}m_j,$ and $\sum _{j=1}^{3}U_{\mu j}^2m_j$. Then, using these 
representations, 
we derive the constraints on the $CP$ violating phases which appear in 
the lepton number violating processes.
\par
This article is organized as follows. In section 2 we introduce the 
graphical representations of 
the complex masses and the $CP$ violating phases.  
In section 3 we 
present constraints on the $CP$ violating phases from $(\beta\beta)_{0\nu}$. 
Constraints from $(\beta\beta)_{0\nu}$ and the neutrino oscillation experiments 
are discussed in section 4. Section 5 is devoted to summary.

\section{graphical representations of the complex masses and $CP$ 
violating phases }
We now rewrite the complex mass $\sum _{j=1}^{3}U_{ej}^2m_j$ by 
using the phase convention in Eq.(\ref{eq772}) as
\begin{eqnarray}
\sum _{j=1}^{3}U_{ej}^2m_j 
& =& c_1^2c_3^2m_1+s_1^2c_3^2e^{2i\beta }m_2+s_3^2e^{2i(\rho -\phi )}m_3 
 \nonumber \\
& =& |U_{e1}|^2m_1+|U_{e2}|^2e^{2i\beta }m_2+|U_{e3}|^2e^{2i(\rho -\phi 
)}m_3  \nonumber \\ 
& \equiv& 
|U_{e1}|^2\widetilde{m_1}+|U_{e2}|^2\widetilde{m_2}+|U_{e3}|^2\widetilde
{m_3}
\label{eq110601}
\end{eqnarray}
Here we have defined the complex masses $\widetilde{m_i} (i=1,2,3)$ by
\begin{mathletters}
\label{eq000131}
\begin{eqnarray}
\widetilde{m_1} & \equiv& m_1  \label{eq2004} \\
\widetilde{m_2} & \equiv& e^{2i\beta }m_2  \label{eq2005} \\
\widetilde{m_3} & \equiv&e^{2i\rho^\prime}m_3,\quad 
\rho^\prime\equiv\rho-\phi.\label{eq2006}
\end{eqnarray}
\end{mathletters}
We also rewrite the complex mass $\sum _{j=1}^{3}U_{\mu j}U_{e j}m_j$ 
by using the above 
$\widetilde{m_i} (i=1,2,3)$ as follows:
\begin{eqnarray}
 \sum _{j=1}^{3}U_{\mu j}U_{e j}m_j
     &=&U_{e1}U_{\mu 1}m_1+U_{e2}U_{\mu2}m_2+U_{e3}U_{\mu3}m_3 
        \nonumber\\
     &=&U_{e1}^*U_{\mu1}\widetilde{m}_1
        +U_{e2}^*U_{\mu2}\widetilde{m}_2
        +U_{e3}^*U_{\mu3}\widetilde{m}_3 \nonumber \\
     &=& U_{e2}^*U_{\mu2}(\widetilde{m}_2-\widetilde{m}_1)
        +U_{e3}^*U_{\mu3}(\widetilde{m}_3-\widetilde{m}_1).
\end{eqnarray}
Here we have used the unitarity constraint that $\sum_{j=1}^3 
U_{ej}^*U_{\mu j}=0$. 
Furthermore, using $U_{\mu1}\equiv|U_{\mu1}|e^{i(\varphi_{21}-\beta)}$, 
$U_{\mu2}\equiv|U_{\mu2}|e^{i\varphi_{22}}$, 
$U_{\mu3}=|U_{\mu3}|e^{i(\rho-\beta )}$, 
$U_{e2}=|U_{e2}|e^{i\beta}$ and 
$U_{e3}=|U_{e3}|e^{i(\rho-\beta)}$  
with $\varphi_{21}\equiv \mbox{arg}{(-s_1c_2-c_1s_2s_3e^{i\phi})}$ 
and $\varphi_{22}\equiv \mbox{arg}{(c_1c_2-s_1s_2s_3e^{i\phi})}$ , 
we obtain
\begin{mathletters}
\label{eq013102}
\begin{eqnarray}
  \sum _{j=1}^{3}U_{\mu j}U_{e j}m_j
   &=& e^{i(\phi-\beta)}
    \left(|U_{e2}^*U_{\mu2}|e^{i(\varphi_{22}-\phi)}
    (\widetilde{m}_2-\widetilde{m}_1)
    +|U_{e3}^*U_{\mu3}|(\widetilde{m}_3-\widetilde{m}_1)\right),\\ 
  \sum _{j=1}^{3}U_{\mu j}^2m_j
   &=&|U_{\mu1}|^2e^{2i(\varphi_{21}-\beta)}m_1
      +|U_{\mu2}|^2e^{2i\varphi_{22}}m_2
      +|U_{\mu3}|^2e^{2i(\rho-\beta)}m_3 \nonumber \\
   &=&e^{2i(\varphi_{21}-\beta)}
    \left(|U_{\mu1}|^2\widetilde{m}_1
      +|U_{\mu2}|^2e^{2i(\varphi_{22}-\varphi_{21})}\widetilde{m}_2
      +|U_{\mu3}|^2e^{2i(\phi-\varphi_{21})}\widetilde{m}_3\right).
\end{eqnarray}
\end{mathletters}
Therefore, the $\langle m_{\nu} \rangle_{e e}$, $\langle m_{\nu} 
\rangle_{\mu e}$, 
and $\langle m_{\nu} \rangle_{\mu \mu}$ defined in Eqs.(\ref{eq761}) 
$\sim$ 
(\ref{eq771}) are reexpressed by the absolute values of averaged complex 
masses as 
\begin{mathletters}
\label{eq013103}
\begin{eqnarray}
\langle m_{\nu} \rangle_{e e}  & =& |M_{ee}|, \\
\langle m_{\nu} \rangle_{\mu e} & =& |M_{\mu e}|, \\
\langle m_{\nu} \rangle_{\mu \mu} & =& |M_{\mu\mu}|.
\end{eqnarray}
\end{mathletters}
Here the averaged complex masses $M_{ee}$, $M_{\mu e}$ ,and $M_{\mu\mu}$ 
are defined by
\begin{eqnarray}
M_{ee}& \equiv& 
|U_{e1}|^2\widetilde{m_1}+|U_{e2}|^2\widetilde{m_2}+|U_{e3}|^2\widetilde
{m_3}, \label{eq2013} \\
M_{\mu e} & \equiv& |U_{e2}^*U_{\mu2}|e^{i(\varphi_{22}-\phi)}
    (\widetilde{m}_2-\widetilde{m}_1)
    +|U_{e3}^*U_{\mu3}|(\widetilde{m}_3-\widetilde{m}_1), \label{eq2014} \\
M_{\mu\mu} & \equiv& |U_{\mu1}|^2\widetilde{m}_1
      +|U_{\mu2}|^2e^{2i(\varphi_{22}-\varphi_{21})}\widetilde{m}_2
      +|U_{\mu3}|^2e^{2i(\phi-\varphi_{21})}\widetilde{m}_3. \label{eq2015}
\end{eqnarray}
\par
Now let us introduce graphical representations of the complex value of 
the 
$M_{ee}$, $M_{\mu e}$ and $M_{\mu\mu}$ in a complex mass plane in 
order to investigate the magnitude of 
the $CP$ violating phases in them.
The $M_{ee}$ is the "averaged" complex mass of the masses 
$\widetilde{m_i} (i=1,2,3)$ weighted by three mixing elements 
$|U_{e j}|^2  (j=1,2,3)$ with the unitarity constraint $\sum 
_{j=1}^{3}|U_{ej}|^2=1$.
Therefore, the position of $M_{ee}$ in a complex mass plane is 
within the triangle formed by the three mass points
$\widetilde{m_i} (i=1,2,3)$ if the magnitudes of $|U_{e j}|^2  (j=1,2,3)
$ are unknown, which is shown in Fig. 1(a-i). 
Hereafter we refer this triangle as the complex-mass triangle.
This triangle is different from that defined 
by Fogli et al. \cite{fogli} in the sense that ours incorporates the $CP$ 
violating phases and masses. 

\par 
The three mixing elements $|U_{e j}|^2  (j=1,2,3)$ indicate 
the division ratios for the three portions of each side of the triangle 
which are divided by 
the parallel lines to the side lines of the triangle passing through the 
$M_{ee}.$ (Fig. 1(a-ii)). 
The $CP$ violating phases $2\beta$ and $2\rho^\prime$ represent the 
rotation angles of 
$\widetilde{m_2}$ and $\widetilde{m_3}$ around the origin,
respectively. 

Likewise, the constraints on the positions of $M_{\mu e}$ and 
$M_{\mu\mu}$ are depicted in Figs.1(b) and 1(c).
The position of $M_{\mu e}$ is given as 
Fig. 1(b).
The position of $M_{\mu\mu}$ in a complex mass plane is 
within the triangle formed by the three mass points
$\widetilde{m_1}$, $e^{2i(\varphi_{22}-\varphi_{21})}\widetilde{m_2}$, 
and $e^{2i(\phi-\varphi_{21})}\widetilde{m}_3$ which is shown in 
Fig. 1(c).

\section{constraints on the $CP$ violating phases from $(\beta\beta)_{0\nu}$}
Among the lepton number violation processes such as $(\beta\beta)_{0\nu}
$, the $\mu^--e^+$ conversion and the K decay, 
$K^-\to\pi^+\mu^-\mu^-$ , the $(\beta\beta)_{0\nu}$ gives us most 
restrictive constraints on the $CP$ violating phases.
Therefore, hereafter we concentrate on the $(\beta\beta)_{0\nu}$ and 
derive constraints on the $CP$ 
violating phases from the experimental upper bound on 
$\langle m_{\nu} \rangle_{ee}$ (we denote it $\langle m_{\nu} 
\rangle_{\mbox{\tiny max}}$, i.e., $\langle m_{\nu} \rangle_{ee} < \langle m_{\nu} 
\rangle_{\mbox{\tiny max}}$).  
Since $\langle m_{\nu} \rangle_{ee}=|M_{ee}|$, 
the present experimental upper bound on $\langle m_{\nu} \rangle_{ee}$ 
obtained from the 
$(\beta\beta)_{0\nu}$ forms the circle in the complex plane and this circle
must include the point \(M_{ee}\) inside of it. 
Namely, the allowed region for \(M_{ee}\) is the intersection of the inside of 
the circle of radius \(\mmmax\) around the origin and the inside of the 
complex-mass triangle which was discussed in section 2.

\par
In the case of \(m_1>\langle m_\nu \rangle_{\mbox{\tiny max}}\), 
we can obtain the constraints on the $CP$ violating phases from 
the allowed region for \(M_{ee}\) without using any constraints on the mixing 
elements \(|U_{ej}|^2  (j=1,2,3)\) as follows. 
In order to obtain the conditions for the allowed \(M_{ee}\), 
it is more convenient to survey the forbidden regions for \(M_{ee}\). 
It is easily understood from Fig 2(a) that the complex-mass triangle
does not overlap with the circle \(\mmmax\) 
only if the following conditions are satisfied for all \(i\) and \(j\). 
\begin{equation}
|\mbox{arg}(\widetilde{m_j}/\widetilde{m_i})|<\alpha_{ij}. \label{eq020301}
\end{equation}
Here \(\alpha_{ij}\) is defined by
\(\alpha_{ij}\equiv \cos^{-1} (\mmmax/m_i)+\cos^{-1}(\mmmax/m_j)\).
Therefore, the allowed region for \(M_{ee}\) is the area where, at least, 
one of the inequalities of Eq. (\ref{eq020301}) is violated.
Since we have 
$|\mbox{arg}(\widetilde{m_2}/\widetilde{m_1})|=|2\beta |$ , 
$|\mbox{arg}(\widetilde{m_3}/\widetilde{m_1})|=|2\rho '|$ , and  
$|\mbox{arg}(\widetilde{m_2}/\widetilde{m_3})|=|2\beta -2\rho '|$, 
with \(2\beta\) and \(2\rho'\) in the interval of \((-\pi,\pi)\),
we find that Majorana $CP$ violating phases $\beta$ and $\rho '$,
must satisfy the following conditions:
\begin{eqnarray}
\alpha_{12}<|2\beta| \quad \mbox{or} \quad 
\alpha_{13}<|2\rho'| \quad \mbox{or} \quad \alpha_{23}<|2\rho'-2\beta|. 
\label{eq110580}
\end{eqnarray}
The allowed region in the \(2\beta\) vs \(2\rho'\) plane obtained from 
Eq.(\ref{eq110580}) is depicted in Fig. 2(b).
\par
Eq.(\ref{eq110580}) is also useful in the case where the three neutrino masses 
are almost degenerate and $\langle m_\nu \rangle_{\mbox{\tiny max}} < m_1\simeq 
m_2\simeq m_3\equiv m$.  In this case, Eq.(\ref{eq110580}) reduces to 
\begin{eqnarray}
\alpha<|2\beta| \quad \mbox{or} \quad 
\alpha<|2\rho'| \quad \mbox{or} \quad \alpha<|2\rho'-2\beta| \label{eq110590} 
\end{eqnarray}
with $\alpha\equiv 2\cos^{-1}(\mmmax/m)$ 
and the allowed region Fig. 2(b) to Fig. 2(c).

\section{constraints on the $CP$ violating phases from $(\beta\beta)_{0\nu}$ 
and the neutrino oscillation experiments}
Now, we consider the constraints on $|U_{ej}|^2$ from the 
CHOOZ reactor experiment, the recent Super-Kamiokande atmospheric 
neutrino experiment, solar neutrino experiments and the future KamLAND 
reactor experiment. 
Then, by combining these constraints on $|U_{ej}|^2$ with one from 
$(\beta\beta)_{0\nu},$ we derive the possible constraints on the $CP$ 
violating 
phases by using our graphical representation. 
In the following discussions we consider three cases for the neutrino 
mass hierarchy, i.e., 
case(A): two quasi-degenerate neutrino with $m_1 \sim m_2 \ll m_3$,  
case(B): two quasi-degenerate neutrino with $m_1 \ll m_2 \sim m_3$ and
case(C): three quasi-degenerate neutrino with $m_1 \simeq m_2 \simeq m_3 
= m$ .

\par
\subsection{two quasi-degenerate neutrino with $m_1 \sim m_2 \ll m_3$ }
In the case, the oscillation probability for 
reactor neutrinos in the three-generation model, $P(\overline{\nu}_e \to 
\overline{\nu}_e)$ is given by 
\begin{eqnarray}
 P(\overline{\nu}_e \to \overline{\nu}_e)
  &=&1-4|U_{e3}|^2(1-|U_{e3}|^2)\sin^2\left(\frac{\Delta m_{13}^2L}{4E}\right)
	\nonumber \\
  &=&1-4s_3^2c_3^2\sin^2\left(\frac{\Delta m_{13}^2L}{4E}\right)
 \label{eq110609}
\end{eqnarray}
if $\Delta m_{13}^2L/(4E)\ll 1.$
The present CHOOZ experiment gives a severe restriction on the mixing 
angle: 
\begin{equation}
 \sin^2 2\theta \alt 0.1.
\end{equation}
In this case, since $\theta=\theta_3,$ we obtain
\begin{equation}
 0 \leq s_3^2 \alt 0.026 \text{\quad or \quad} 0.97 \alt s_3^2 
\leq 1.
 \label{eq111}
\end{equation}
\par
On the other hand, the oscillation probability for the atmospheric 
neutrinos, 
 $P(\nu_\mu \to \nu_\mu)$ is
\begin{eqnarray}
 P(\nu_\mu \to \nu_\mu)
  &=&1-4|U_{\mu3}|^2(1-|U_{\mu3}|^2)\sin^2\left(\frac{\Delta m_{13}^2L}{4E}\right)
	\nonumber \\
  &=&1-4c_3^2s_2^2(1-c_3^2s_2^2)\sin^2\left(\frac{\Delta 
m_{13}^2L}{4E}\right).
  \label{eq110608}
\end{eqnarray}
The atmospheric $\nu_\mu$ deficit in the Super Kamiokande experiment 
indicates 
that $0.8 \alt 4c_3^2s_2^2(1-c_3^2s_2^2) \leq 1,$ namely we obtain 
\begin{equation}
 \frac{0.28}{1-s_3^2} \alt s_2^2 \alt \frac{0.72}{1-s_3^2}.
 \label{eq222}
\end{equation}
From these two constraints, Eqs.(\ref{eq111}) and (\ref{eq222}), we 
obtain 
\begin{equation}
 0.28 \alt s_2^2 \alt 0.74,\qquad s_3^2 \alt 0.026.
 \label{eq110604}
\end{equation}
Eq.(\ref{eq110604}) imposes the restriction on $|U_{e3}|^2$. 
Therefore, when combined with the allowed region in the complex mass 
plane discussed in the section 2, 
the position of $M_{ee}$ in our graphical representation is  
restricted by the CHOOZ and Super Kamiokande experiments as shown in 
Fig. 3.
We also have the constraints on the mixing angle from the solar neutrino 
experiments. They give several 
separate allowed regions for the position of $M_{ee}$ in our 
graphical representation as shown in Fig. 3. 
Whether the mixing angle for solar neutrinos 
is large or small can be determined by the future KamLAND reactor 
experiment \cite{kamland}. 
The future KamLAND experiment will also lead to the constraint on the 
$|U_{e1}|^2$ and $|U_{e2}|^2$. Since the KamLAND experiment has the 
chance to 
observe a lower order mass difference, 
$\Delta m^2 \sim 10^{-5} \text{eV}^2,$ we can't neglect the term depend 
on 
$\Delta m_{12}^2 \text{\, in \,} P(\overline{\nu}_e \to 
\overline{\nu}_e).$ 
So we rewrite Eq.(\ref{eq110609}) as follows:
\begin{eqnarray}
 P(\overline{\nu}_e \to \overline{\nu}_e)
  &=&1-4 \left[\frac{|U_{e3}|^2(1-|U_{e3}|^2)}{2}+
        |U_{e1}|^2|U_{e2}|^2\sin^2\left(\frac{\Delta m_{12}^2L}{4E}\right)\right]
	\nonumber \\
  &=&1-\left[\frac{2s_3^2c_3^2}{\sin^2\left(\frac{\Delta m_{12}^2L}{4E}\right)}
       +4 s_1^2c_1^2c_3^4\right]
	\sin^2\left(\frac{\Delta m_{12}^2L}{4E}\right).\nonumber \\
  &\equiv& 1-\Xi^2
	\sin^2\left(\frac{\Delta m_{12}^2L}{4E}\right).
\label{eq020701}
\end{eqnarray}
Here we have used the following conditions,
\begin{equation}
 \sin^2\left(\frac{\Delta m_{13}^2L}{4E}\right) \sim \frac{1}{2},\qquad
 \sin^2\left(\frac{\Delta m_{23}^2L}{4E}\right) \sim \frac{1}{2},
\end{equation}
because of their frequent oscillations. 
Let us combine Eq.(\ref{eq020701}) with 
the constraint given in  Eq.(\ref{eq110604}) which is obtained from 
the CHOOZ and Super Kamiokande experiments.  
Then the KamLAND experiment will give the constraints on $|U_{e1}|^2$ and 
$|U_{e2}|^2$,  which will restrict the allowed region for the position of 
$M_{ee}$ as shown in Fig. 4. 
\par
Now, with use of our graphical representation, 
we proceed to discuss the main subject in this paper: 
If we have non zero value of $\langle m_\nu \rangle_{ee}$, 
how can we determine the magnitude of Majorana $CP$ phases 
$\beta \text{\,or\,} \rho^\prime$ ? 
\par
First we discuss the simple case in which $|U_{e3}|^2$ is approximately 
zero and
 the large mixing angle solution(LMA), 
 $0.2 \alt |U_{e1}|^2 \alt 0.8$, is adopted for the solar 
neutrino problem. 
In this case, we have
\begin{eqnarray}
\langle m_\nu \rangle_{ee} &\simeq& \mid |U_{e1}|^2\widetilde{m}_1
                                 +|U_{e2}|^2\widetilde{m}_2 \mid 
\nonumber\\
           &\equiv& \mid |U_{e1}|^2\widetilde{m}_1
                                 +(1-|U_{e1}|^2)\widetilde{m}_2 \mid.
           \label{eq110605}
\end{eqnarray}
Given the values of $\langle m_\nu \rangle_{ee},m_1,m_2,
|U_{e1}|^2 \text{\,and }|U_{e2}|^2$, 
the $CP$ violating phase $\beta$ is easily obtained from the 
graphical representation of Eq.(\ref{eq110605}). 
It goes from Fig.1(a-ii) that the complex-mass triangle 
gets degenerate to a straight line \(\widetilde{m_1}\widetilde{m_2}\) 
for \(|U_{e3}|^2=0\) case and that the position of 
$M_{ee}=|U_{e1}|^2\widetilde{m_1}+|U_{e2}|^2\widetilde{m_2}$ moves along the circle 
with a radius of $|U_{e2}|^2m_2(=(1-|U_{e1}|^2)m_2)$ from the point \(|U_{e1}|^2 m_1\) for changing $\beta$.
On the other hand, the measurement of the $\langle m_\nu \rangle_{ee}$ restricts 
$M_{ee}$ on the circle with a radius of $\langle m_\nu \rangle_{ee}$ from the origin. 
Therefore, the $\beta$ is determined by the intersection of the above two circles as shown in Fig.5. Applying the cosine formula to \(\triangle OAB\) in Fig.5,
we find 
\begin{equation}
\mm^2 = |U_{e1}|^4 m_1^2+|U_{e2}|^4 m_2^2
	+2|U_{e1}|^2|U_{e2}|^2 m_1 m_2 \cos2\beta.
\end{equation}
Therefore, we obtain 
\begin{equation}
\cos 2\beta=\frac{\mm^2 - |U_{e1}|^4 m_1^2-|U_{e2}|^4 m_2^2}
		{2|U_{e1}|^2|U_{e2}|^2 m_1 m_2}. \label{eq110616}
\end{equation}
It goes from Eq.(\ref{eq110616}) with the use of $-1\le\cos 2\beta\le1$ that \(\mm_{ee}\) has the lower and 
upper limits as $\mm_{\mbox{\tiny lower}}\le\mm_{ee}\le\mm_{\mbox{\tiny upper}}$,  which is shown in Fig.6 with the definitions of 
\begin{eqnarray}
\mm_{\mbox{\tiny lower}}=|m_1-|U_{e2}|^2(m_1+m_2)| \nonumber\\
\mm_{\mbox{\tiny upper}}=m_1+|U_{e2}|^2(m_2-m_1).\label{eq110626}
\end{eqnarray}
On the other hand, for the case where 
$|U_{e3}|^2 \simeq 0$ and the small mixing angle solution(SMA) 
is adopted for the solar neutrino problem, i.e., 
$\theta_1=0\text{\, or \,}\pi/2,$ 
we can not obtain any information about $\beta$, 
since we have $\langle m_\nu \rangle_{ee}=|m_1|=m_1$ for 
$\theta_1=0$ or 
$\langle m_\nu \rangle_{ee}=\mid e^{2i\beta}m_2 \mid=m_2$ for 
$\theta_1=\pi/2$.
\par
Second we consider the case where $|U_{e3}|^2\ne0$ and the LMA solution, 
 $0.2 \alt |U_{e1}|^2 \alt 0.8$, is adopted for the solar neutrino problem.
In this case, we have
\begin{equation}
M_{ee}-|U_{e3}| \widetilde{m_3}=|U_{e1}|^2m_1+|U_{e2}|^2 \widetilde{m_2}.
\label{eq022201}
\end{equation}
The graphical representation of Eq.(\ref{eq022201}) 
is shown in Fig.7. 
In Fig.7(a) we consider the case in which the circle of radius \(|U_{e2}|^2 m_2\) 
around the point \((|U_{e1}|^2 m_1,0)\) (which we refer as \(A\) or 
\(\overrightarrow{OA}\)) intersects 
with the circles of radius \(\mm_{ee}\pm |U_{e3}|^2m_3\) around the origin  
at the points \(B_1\) and \(B_2\). 
We find that \(2\beta\) is ranging from the argument of 
\(\overrightarrow{AB_1}\) to that of 
\(\overrightarrow{AB_2}\) as seen in Fig.7(a). 
The relation between \(\beta\) and \(\rho'\) is also derived 
from Eq.(\ref{eq022201}): 
For fixed \(2\beta\), the \(\rho'\) has two solution \(\rho'_1\) and \(\rho'_2\)  
which are determined by the points \(C_1\) and \(C_2\)  as shown in 
Fig.7(b). 
Here the \(C_1\) and \(C_2\) are the intersections of the circle of 
radius \(|U_{e3}|^2 m_3\) around 
the point $|U_{e1}|^2\widetilde{m_1}+|U_{e2}|^2\widetilde{m_2}$ (which we refer as \(B\)) 
with the circle of radius \(\mm_{ee}\) around the origin 
since \(\overrightarrow{OA}+\overrightarrow{AB}+\overrightarrow{BC}=M_{ee}\)
from Eq.(\ref{eq022201}). 
Thus we obtain the relation between \(\beta\) and \(\rho'\). 
We depict this relation in Fig. 8. 
The other cases may occur but they can be treated analogously.

\subsection{two quasi-degenerate neutrino with $m_1 \ll m_2 \sim m_3$ }
In this case, the CHOOZ experiment and the atmospheric neutrino deficit 
experiment indicates $|U_{e1}|^2\sim0$ as is seen in 
Fig. 9.
Therefore, we can discuss this case with the same way as the case(A) only 
by replacing $m_1$, $|U_{e1}|^2$, and $\beta$ with $m_3$, $|U_{e3}|^2$, 
and $\beta-\rho'$, respectively.

\par
\subsection{three quasi-degenerate neutrino with $m_1 \simeq m_2 \simeq 
m_3 = m$}
We assume that all three neutrino masses are almost degenerate, then we 
have 
\begin{equation}
 M_{ee}=m\left(|U_{e1}|^2+|U_{e2}|^2e^{2i\beta}
           +|U_{e3}|^2e^{2i\rho^\prime}\right). \nonumber 
\end{equation}
The constraints on the $CP$ violating phases from this $M_{ee}$
is obtained from the similar discussions as in Fig. 7 with only taking $m_1=m_2=m_3=m$ in it. 
It should be noted that for the case where $\Delta m_{12}^2\ll\Delta m_{13}^2$ and $|U_{e3}|^2$ is 
approximately zero, we find 
\begin{eqnarray}
 \sin^2\beta&=&\frac{m^2-\langle m_\nu \rangle_{ee}^2}
                    {4|U_{e1}|^2(1-|U_{e1}|^2)m^2} \nonumber \\
         &=&\frac{m^2-\langle m_\nu \rangle_{ee}^2}
                 {4s_1^2(1-s_1^2)m^2}.   
\label{eq555}
\end{eqnarray}
which is the same result as one obtained from Eq.(\ref{eq110616}) with replacing $m_{i}(i=1,2,3)$ with $m$. 
We find from Eq.(\ref{eq555}) that the following lower limit of $\sin^2{\beta}$ is obtained for 
the large mixing angle solution(LMA), $0.2 \alt |U_{e1}|^2=s_1^2 \alt 0.8$, of the solar neutrino 
problem, 
\begin{equation}
\sin^2{\beta} \ge 1-\left(\frac{\langle m_\nu \rangle_{ee}}{m}\right)^2, 
\end{equation}
where the lower limit is realized at $s_1^2=0.5$.

\section{summary }
We have introduced graphical representations of the complex masses, 
$M_{ee}$, $M_{\mu e}$ and $M_{\mu\mu}$ whose absolute magnitudes are 
experimentally observable "averaged" masses,  
$\langle m_{\nu} \rangle_{e e}$, $\langle m_{\nu} \rangle_{\mu e}$ 
and $\langle m_{\nu} \rangle_{\mu \mu}$ 
of the lepton number violation processes such as neutrinoless double 
beta decay, the $\mu^--e^+$ conversion and the K decay, 
$K^-\to\pi^+\mu^-\mu^-$ , respectively. By using those graphical 
representations, 
we have investigated how to determine the magnitude of the $CP$ 
violating phases from the analysis of the neutrinoless double 
beta decay. 
First we have discussed without using any constraint on the mixing elements 
$|U_{e j}|^2  (j=1,2,3)$ and obtained 
the constraints on the Majorana $CP$ violating phases, 
Eqs.(\ref{eq110580}) and (\ref{eq110590})  if $\langle m_{\nu} 
\rangle_{\mbox{\tiny max}} < m_1$, from which 
the allowed region in the $2\beta \mbox{\, vs \,} 2\rho^\prime$ plane 
have been derived and shown in Fig. 2.
Of course, we have no constraint on the Majorana $CP$ phases if $\langle 
m_{\nu} \rangle_{\mbox{\tiny max}} > m_1$. 
Still, Eq.(\ref{eq110590}) is useful 
if the three neutrino masses are almost degenerate and 
$\langle m_{\nu} \rangle_{\mbox{\tiny max}} < m_1 \simeq m_2 \simeq m_3 = m$. 
Next by using the constraints on the mixing elements $|U_{e j}|^2  (j=1,
2,3)$ 
obtained from the recent Super Kamiokande atmospheric neutrino 
experiment, 
the solar neutrino experiment, the recent CHOOZ reactor 
experiment, and the future KamLAND reactor 
experiment, we have further discussed the possible constraints on the Majorana 
$CP$ violating phases 
for three cases for the neutrino mass hierarchy, i.e.,
case(A): two quasi-degenerate neutrino with $m_1 \sim m_2 \ll m_3$,  
case(B): two quasi-degenerate neutrino with $m_1 \ll m_2 \sim m_3$ and
case(C): three quasi-degenerate neutrino with $m_1 \simeq m_2 \simeq m_3 
= m$ .
In the case(A), we have obtained the expression of 
$\cos2{\beta}$, Eq.(\ref{eq110616}), in terms of $m_1,m_2, \langle 
m_{\nu} \rangle_{ee}$, $|U_{e1}|^2$, and $|U_{e2}|^2$
for the simple case where $|U_{e3}|^2\simeq0$ with the use of the large mixing angle 
solution(LMA) for the solar neutrino problem. 
The $\langle m_\nu \rangle_{ee}\text{\,\,vs\,\,}\cos2\beta$ relation is shown 
in Fig. 5. Using $-1\le\cos2{\beta} \le 1$, we have found that 
the $\langle m_\nu \rangle_{ee}$ has the lower and upper limits as given 
in Eq.(\ref{eq110626}).
We have also obtained the relation 
between $2\beta \mbox{\, and \,} 2\rho^\prime$ for $|U_{e3}|^2\ne0$ case 
which is shown in Figs. 7 and 8.
We can discuss the case(B) using the same way as the case(A)
by replacing $m_1$, $|U_{e1}|^2$, and $\beta$ with $m_3$, $|U_{e3}|^2$, 
and $\beta-\rho'$, respectively. 
In the case(C), we have obtained the expression of 
$\sin^2{\beta}$ given in Eq.(\ref{eq555}) 
for the simple case where $\Delta m_{12}^2\ll\Delta m_{13}^2$ and 
$|U_{e3}|^2\simeq0$ with the use of the LMA solution for the solar 
neutrino problem. 
From this relation 
we have found that the lower limit of 
$\sin^2{\beta}$ is given by  
$\sin^2{\beta} \ge 1-(\langle m_\nu \rangle_{ee}/m)^2$ for the LMA solution. 

\ \\
Acknowledgement

We are greatly indebted to O.Yasuda for useful discussions.


\newpage
\begin{center}
\includegraphics{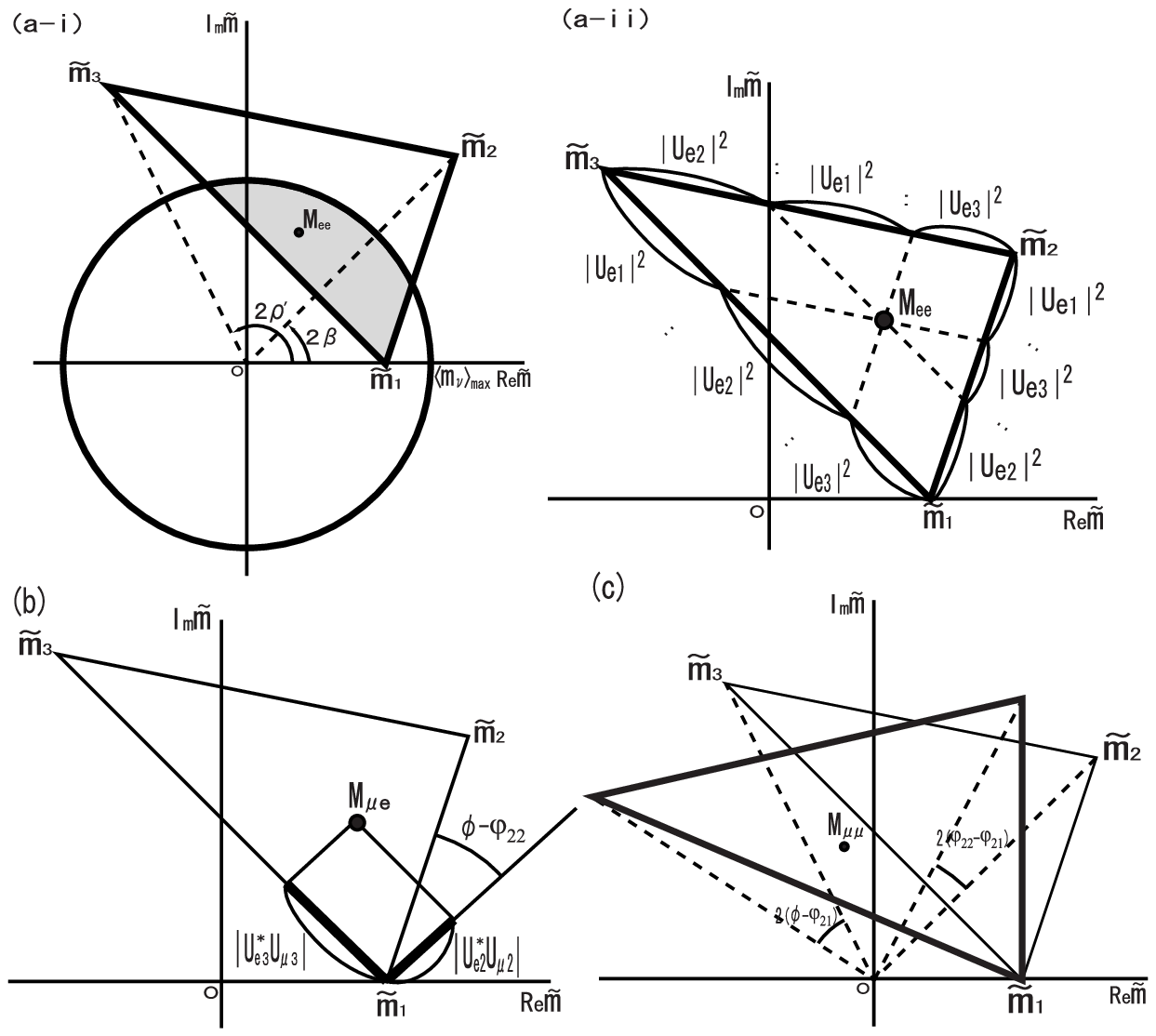}
\end{center}
FIG.1 Graphical representations of the $CP$ violating phases and the complex masses 
$M_{ee}$, $M_{\mu e}$ and $M_{\mu\mu}$ defined in Eqs.(\ref{eq2013})-(\ref{eq2015}).
(a-i) The complex-mass triangle for $M_{ee}$
is formed by the three points $\widetilde{m_i} (i=1,2,3)$ 
defined in Eqs.(\ref{eq2004})-(\ref{eq2006}). 
The allowed position of \(M_{ee}\) is in the intersection (shaded area)
of the inside of this triangle and the inside of the circle of 
radius \(\mmmax\) around the origin.
(a-ii) The relations between the position of $M_{ee}$ and $U_{ei} (i=1,2,
3)$ components of MNS mixing matrix.
(b)  The position of $M_{\mu e}$. 
The position of $M_{\mu e}$ is at the vertex of the parallelogram
of which the other vertexes are at 
$\widetilde{m_1}$, 
$|U_{e2}^*U_{\mu 2}|e^{2i(\varphi_{22}-\varphi_{21})}(\widetilde{m_2}-\widetilde{m_1})$, 
and $|U_{e3}^*U_{\mu 3}|e^{2i(\phi-\varphi_{21})}(\widetilde{m}_3-\widetilde{m_1})$. 
Namely, 
rotate $\widetilde{m_2}$ clockwise by $\phi-\varphi_{22}$ around 
$\widetilde{m_1}$ and scale down by $|U_{e2}^*U_{\mu 2}|$.  From this 
point extend the line parallel to the side of 
$\widetilde{m_1}\widetilde{m_3}$ by $|U_{e3}^*U_{\mu 3}||\widetilde{m_3}
-\widetilde{m_1}|$, 
then we obtain the position of $M_{\mu e}$.
(c) The complex-mass triangle for $M_{\mu\mu}$ (thick lines).
The allowed position of $M_{\mu\mu}$ is within the triangle formed by 
the three points $\widetilde{m_1}$, $e^{2i(\varphi_{22}-\varphi_{21})}\widetilde{m_2}$, 
and $e^{2i(\phi-\varphi_{21})}\widetilde{m}_3$.

\begin{center}
\includegraphics{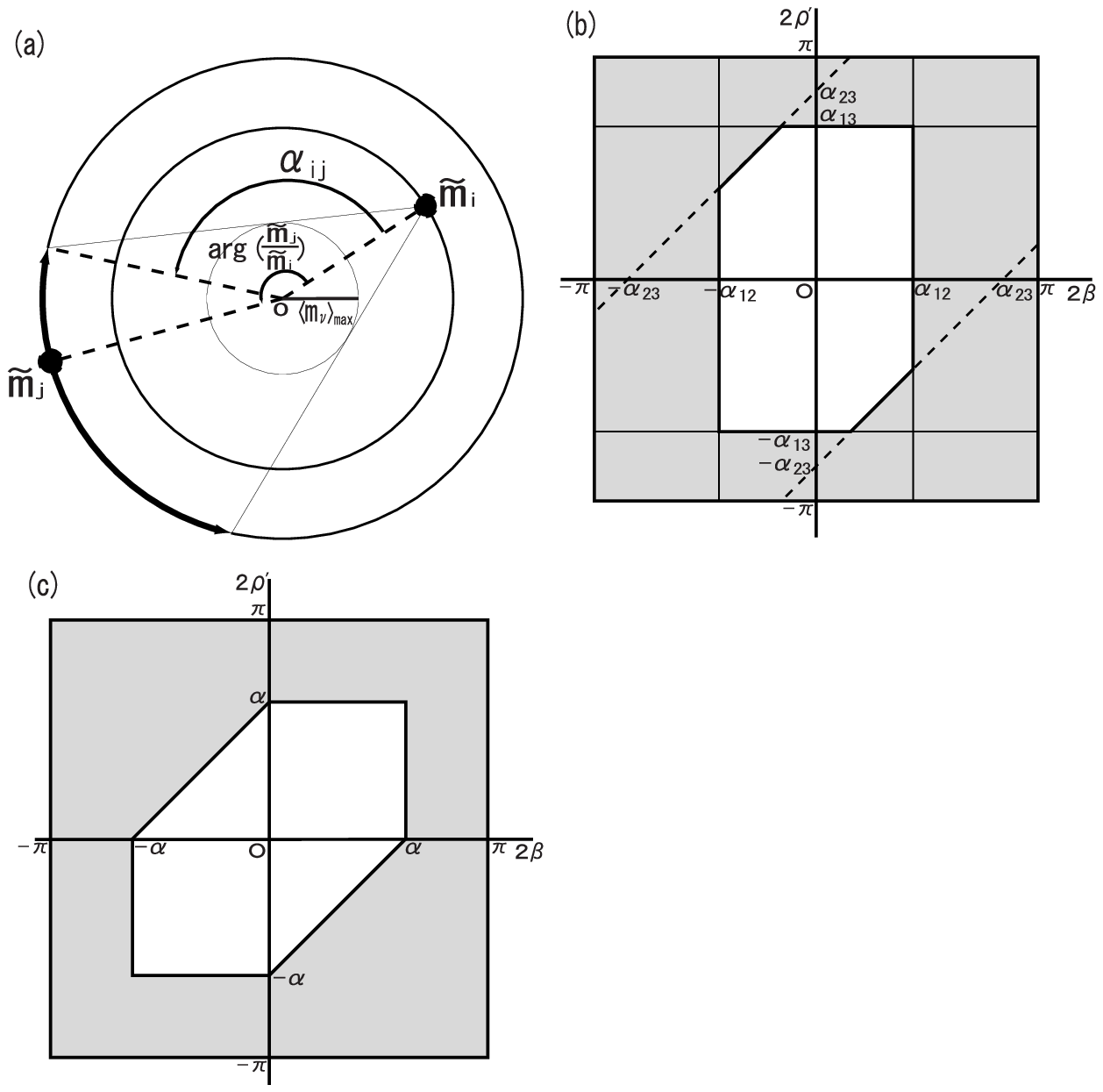}
\end{center}
FIG.2 The restrictions of CP violating phases $2\beta$ and $2\rho '$ from
$(\beta\beta)_{0\nu}$ with arguments independent of $|U_{ej}|^2$.
(a) The allowed region of $M_{ee}$ is inside of the complex-mass triangle
 overlapped with the inside of the circle of radius 
$\langle m_{\nu} \rangle_{\mbox{\tiny max}}$. 
The case where 
the conditions \(|\alpha _{ij}| > |\mbox{arg}(\widetilde{m_j}/\widetilde{m_i})|\) are 
satisfied for all \(i\) and \(j\) is excluded 
since the triangle and the circle can not overlap each other. 
Here we define \(\alpha_{ij}\equiv \cos^{-1}(\mmmax/m_i)+\cos^{-1}(\mmmax/m_j)\), 
$|\mbox{arg}(\widetilde{m_2}/\widetilde{m_1})|\equiv |2\beta |<\pi$ , 
$|\mbox{arg}(\widetilde{m_3}/\widetilde{m_1})|\equiv |2\rho '|<\pi$ , and, therefore, 
$|\mbox{arg}(\widetilde{m_2}/\widetilde{m_3})|\equiv |2\beta -2\rho '|<2\pi$. 
(b) The allowed region (shaded area) 
in the \(2\beta\) vs \(2\rho'\) plane for 
\(\langle m_{\nu} \rangle_{\mbox{\tiny max}} <m_1<m_2<m_3\) case.
(c) The special case of Fig.2 (b) for the case in which three neutrinos 
have almost degenerate masses with 
\(\langle m_\nu \rangle_{\mbox{\tiny max}} <m_1\simeq m_2\simeq m_3\). 
Here \(\alpha\equiv 2\cos^{-1}(\mmmax/m)\).

\begin{center}
\includegraphics{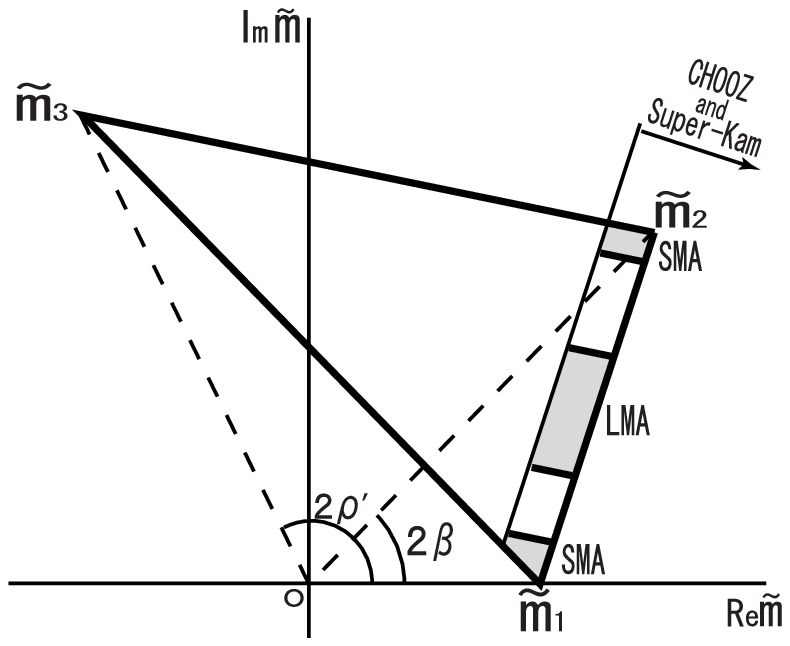}
\end{center}
FIG.3 The allowed region (shaded area) 
of $M_{ee}$ for the case (A) from the CHOOZ, 
the atmospheric \(\nu_\mu\) deficit and the solar neutrino experiments. 
The CHOOZ and the atmospheric \(\nu_\mu\) deficit experiments restrict 
the position of $M_{ee}$ in 
the neighborhood of the side $\widetilde{m_1}\widetilde{m_2}$ . 
The large mixing angle (LMA) and 
the small mixing angle (SMA) solutions for the solar neutrino problem give 
separate allowed regions.

\newpage
\begin{center}
\includegraphics{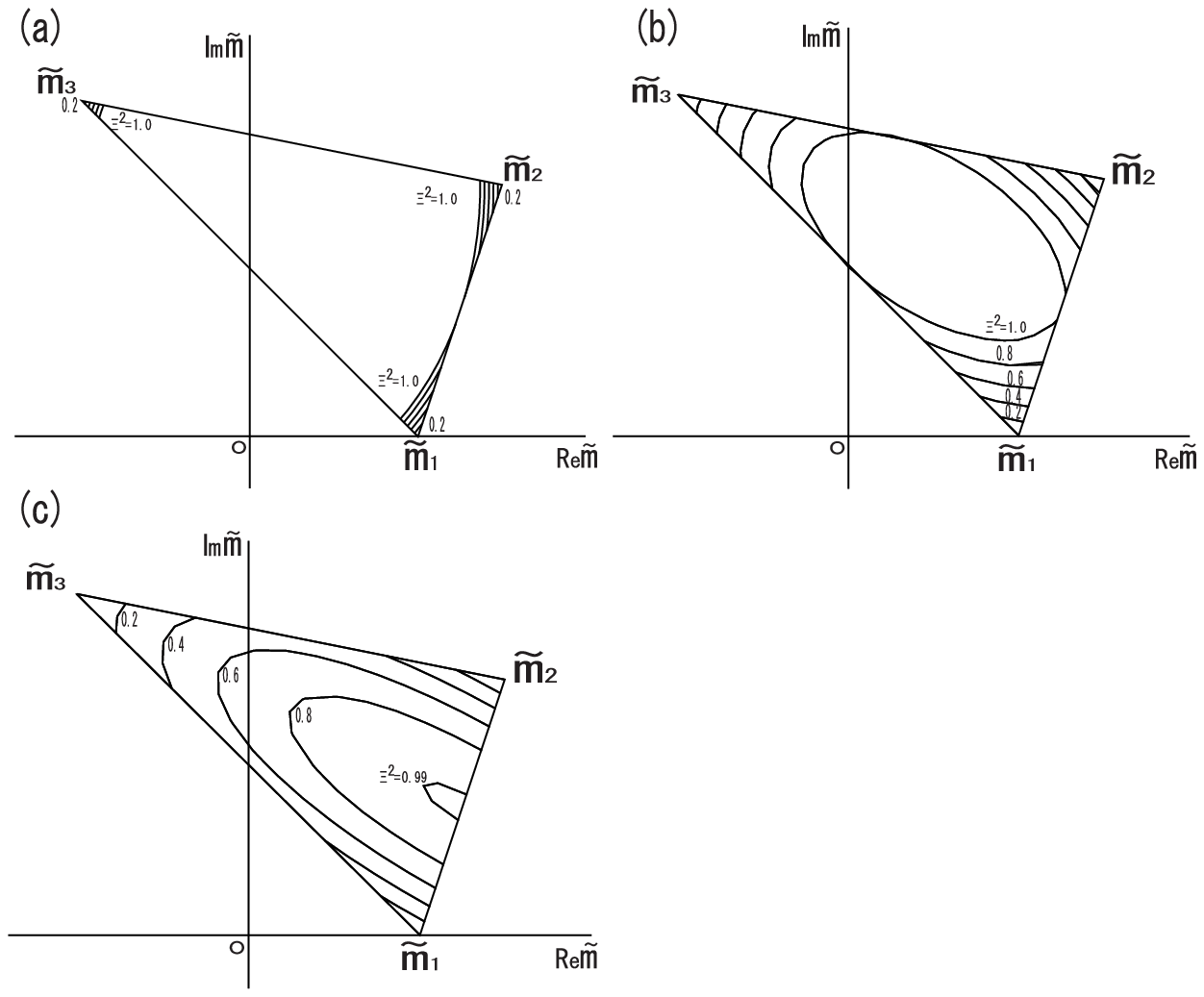}
\end{center}
FIG.4 Constraints from KamLAND experiments. 
The contours of P$(\overline{\nu_e}\rightarrow \overline{\nu_e})$ in our complex-mass triangle. (\(\Xi\) is defined in Eq.(4.7).) 
They are plotted at the interval of $0.2$ of 
$\Xi^2$ for the typical values of $\sin^2\frac{\Delta m^2_{12}}{4E}L$. 
(a); $\sin^2\frac{\Delta m^2_{12}}{4E}L=0.1$.
(b); $\sin^2\frac{\Delta m^2_{12}}{4E}L=0.5$.
(c); $\sin^2\frac{\Delta m^2_{12}}{4E}L=1$.

\begin{center}
\includegraphics{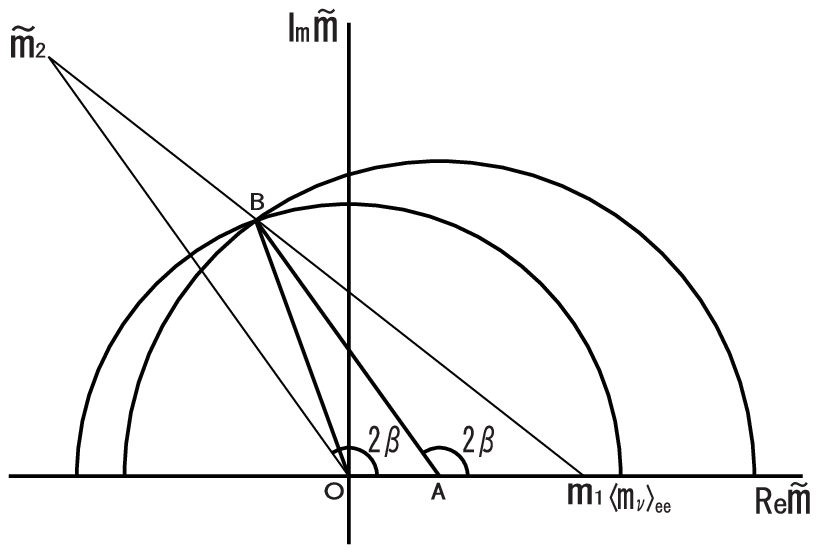}
\end{center}
FIG.5 The determination of \(\beta\) for \(|U_{e3}|^2=0\) in the case(A).
The \(\beta\) is determined from the point \(B\) which is the intersection of 
two circles; the circle of radius \(\mm_{ee}\) around the origin and 
that of radius of \(|U_{e2}|^2m_2\) around \((|U_{e1}|^2 m_1,0)\) (which we refer as \(A\)).
The line \(AB\) is parallel to the line \(O \widetilde{m_2}\).  
Here \(|OA|=|U_{e1}|^2 m_1\), \(|OB|=\mm_{ee}\), \(|AB|=|U_{e2}|^2 m_2\).

\begin{center}
\includegraphics{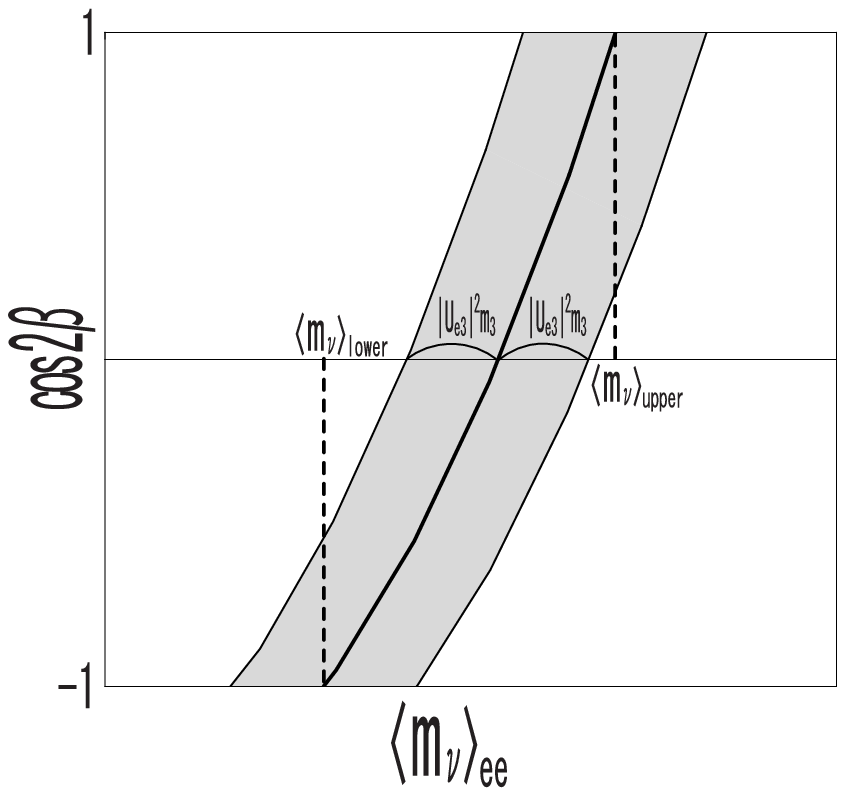}
\end{center}
FIG.6 The relation between \(\cos 2\beta\) and \(\mm_{ee}\) 
which is obtained from Eq.(4.11). 
The solid line is for the case, \(U_{e3}=0\). For \(U_{e3}\ne0\) case,
the relation has a band structure (shaded region).
Here we define \(\mm_{\mbox{\tiny lower}}
\equiv |m_1-|U_{e2}|^2(m_1+m_2)|\) and 
\(\mm_{\mbox{\tiny upper}}\equiv m_1+|U_{e2}|^2(m_2-m_1)\).

\begin{center}
\includegraphics[width=10cm]{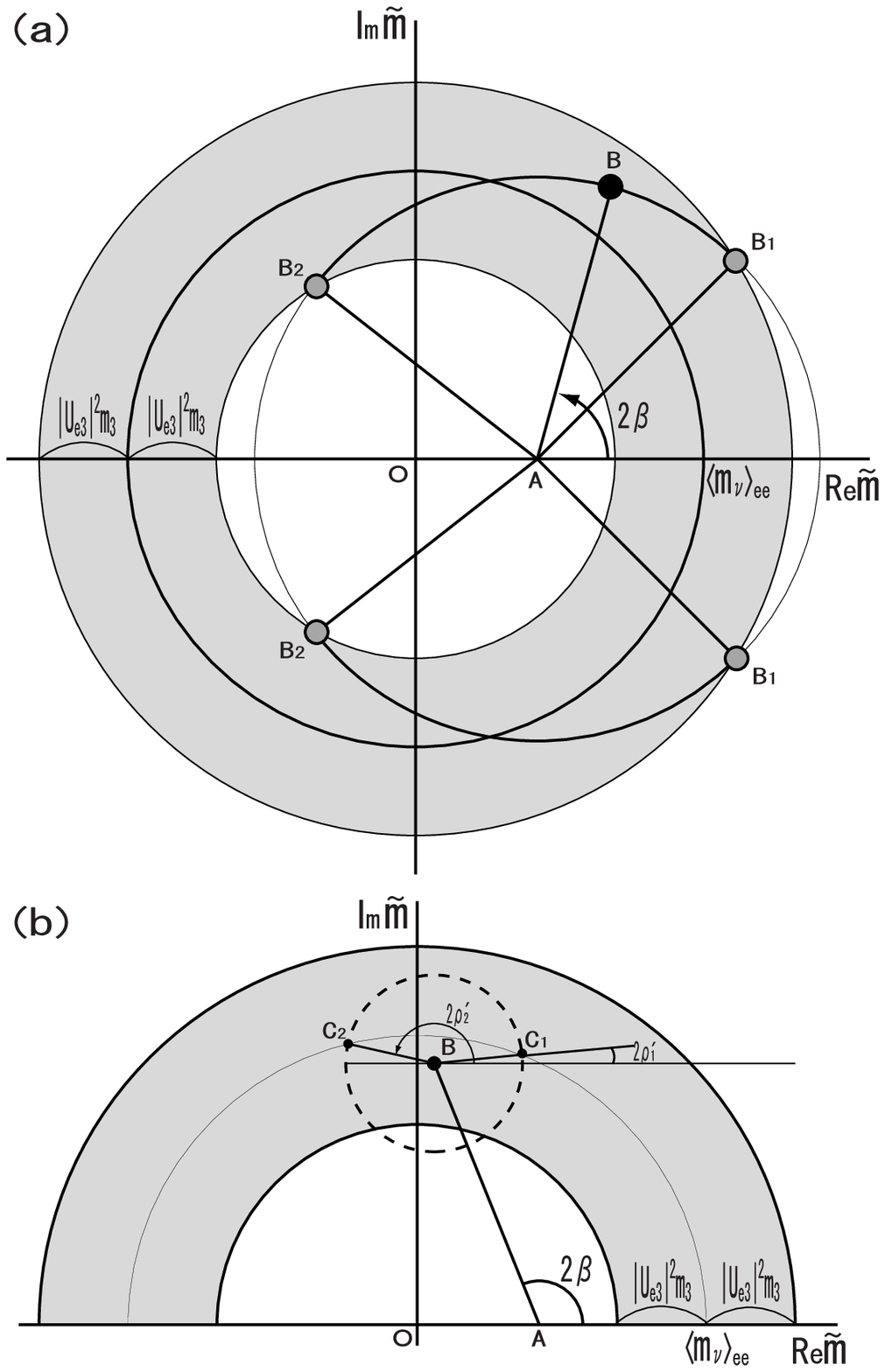}
\end{center}
FIG.7 The constraint on \(\beta\) and \(\rho'\) 
for \(|U_{e3}|^2\neq 0\) in the case(A).
(a) The \(2\beta\) is ranging from the argument of \(\overrightarrow{AB_1}\) 
(angle between \(\overrightarrow{AB_1}\) and the horizontal axis)
to that of \(\overrightarrow{AB_2}\). 
Here \(OA=|U_{e1}|^2 m_1\) and \(|AB|=|U_{e2}|^2 m_2\). 
In this diagram we consider the case where the circle 
of radius \(|U_{e2}|^2 m_2\) 
around \(A\) intersects with the circles of radius 
\(\mm_{ee}\pm |U_{e3}|^2m_3\) at \(B_1\) and \(B_2\). 
(b) The \(\rho'\) has two solution \(\rho'_1\) and \(\rho'_2\) 
for fixed \(2\beta\) since 
\(\overrightarrow{OA}+\overrightarrow{AB}+\overrightarrow{BC}=M_{ee}\).
The dotted line is the circle of radius \(|U_{e3}|^2 m_3\) around \(B\) 
which intersects the circle of radius \(\mm_{ee}\) around the origin at \(C_1\) and \(C_2\). 
Here we refer the point $|U_{e1}|^2\widetilde{m_1}+|U_{e2}|^2\widetilde{m_2}$ as \(B\).

\begin{center}
\includegraphics{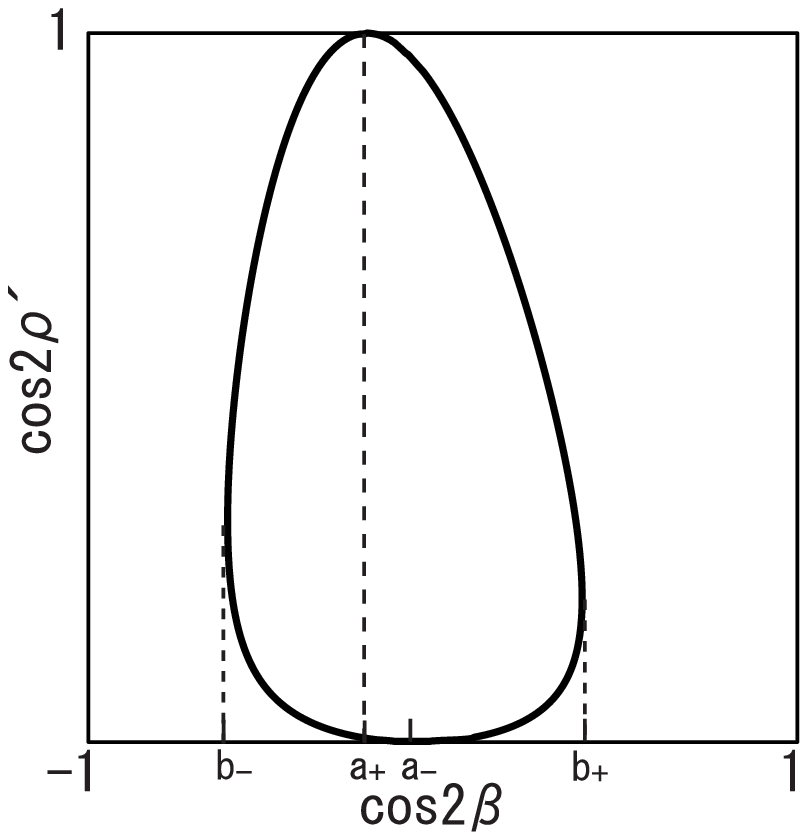}
\end{center}
FIG.8 The relation between \(\cos2\beta\) and \(\cos2\rho'\).
The \(2\rho'\) has always two solutions for fixed \(\beta\) as shown in Fig. 7(b). 
The \(a_{\pm}\) and \(b_{\pm}\) are given 
by \(a_{\pm}\equiv
 \frac{\mm_{ee}^2-(|U_{e1}|^2m_1\pm|U_{e3}|^2m_3)^2-|U_{e2}|^4m_2^2}
{2|U_{e2}|^2m_2(|U_{e1}|^2m_1\pm|U_{e3}|^2m_3)}\) 
and \(b_{\pm}\equiv
 \frac{(\mm_{ee}\pm|U_{e3}|^2m_3)^2 - |U_{e1}|^4m_1^2-|U_{e2}|^4m_2^2}
{2|U_{e1}|^2|U_{e2}|^2m_1 m_2}\).

\begin{center}
\includegraphics{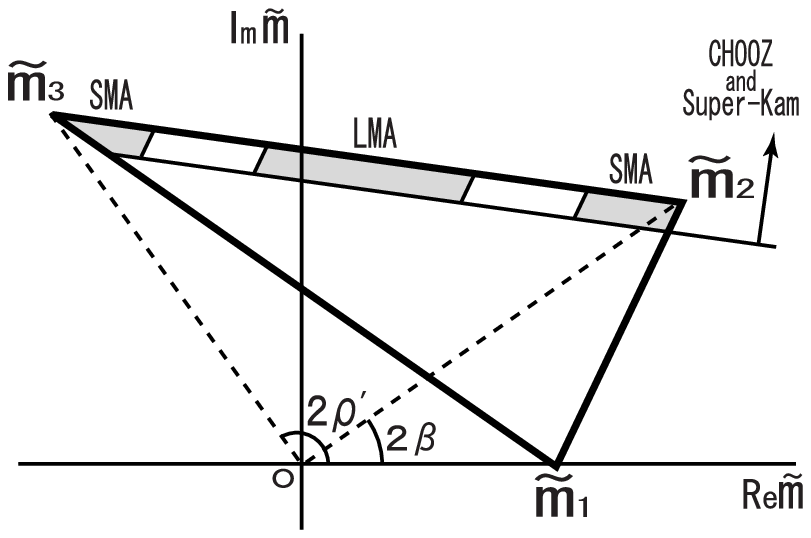}
\end{center}
FIG.9 The allowed region of $M_{ee}$ for the case (B) 
from the CHOOZ and the Super Kamiokande experiments. 
The position of $M_{ee}$ is restricted in the shaded area which is in 
the neighborhood of the edge $\widetilde{m_2}\widetilde{m_3}$.

\end{document}